\documentclass{aastex631}

\hypersetup{linkcolor=red,citecolor=blue,filecolor=cyan,urlcolor=magenta}

\submitjournal{RNAAS}

\shorttitle{1RXS J064434.5+334451}
\shortauthors{Shafter \& Bautista}

\graphicspath{{./}{figures/}}

\begin{document}

\title{Updated Ephemeris and Evidence for a Period Change in the Eclipsing Novalike Variable 1RXS J064434.5+334451}

\correspondingauthor{A. W. Shafter}
\email{ashafter@sdsu.edu}

\author[0000-0002-1276-1486]{A. W. Shafter}
\affiliation{Department of Astronomy, San Diego State University}

\author[0000-0002-1450-8644]{Vladimir Bautista}
\affiliation{Department of Astronomy, San Diego State University}

\begin{abstract}

We report seven new eclipse timings for the novalike variable
1RXS J064434.5+334451.
An analysis
of our data, along with all previously available timings
(36 published and 16 unpublished),
yields a best-fitting linear ephemeris of
BJD$_\mathrm{ecl} = 2,453,403.7611(2) + 0.269~374~43(2)~\mathrm{E}$.
We find a somewhat improved fit with a quadratic ephemeris given by:
BJD$_\mathrm{ecl} = 2,453,403.7598 + 0.269~374~87~\mathrm{E} - 2.0\times10^{-11}~\mathrm{E}^2$,
which suggests that the orbital period may be decreasing at a rate given by
$\dot P \simeq -1.5\times10^{-10}$.

\end{abstract}


\section{Introduction} \label{sec:intro}

The bright ($V\sim13.5$) novalike variable 
1RXS J064434.5+334451 (hereafter J0644)
was discovered to be an eclipsing binary
by \citet{2007A&A...474..951S}. Their photometric
observations over 16 nights between 2005 Feb 02 UT and 2006 Oct 13 UT
revealed deep ($1-1.2$ mag) eclipses that recurred with an
orbital period, $P_\mathrm{orb} = 6.4649808\pm0.0000060$~hr.
Five years later
\citet{2012JAVSO..40..295B} used 22 new eclipse timings, along
with 20 unpublished eclipse timings
from 2005 to 2008 (made available
by D. Sing and B. Green from their original study)
to refine the period and determine an eclipse ephemeris
given by:
HJD$_\mathrm{ecl} = 2453403.75955(12) + 0.26937447(4)~\mathrm{E}$.
Unfortunately, the 20 unpublished eclipse timings remained unavailable,
as they were not reproduced in the \citet{2012JAVSO..40..295B} paper.

The most recent determination of the eclipse ephemeris was made by
\citet{2017MNRAS.464..104H}. These authors acquired the original data
for four eclipses from the work of \citet{2007A&A...474..951S} and remeasured
the times of mid-eclipse. They then combined these timings
with a total of 10
additional timings that were obtained from eclipses
observed in 2008 January and
2010 November and December finding
HJD$_\mathrm{ecl} = 2,453,403.759533 + 0.269~374~46~\mathrm{E}$,
which is nearly identical to the earlier ephemeris. The similarity
is remarkable given that
these authors were apparently unaware of the
22 timings that were reported in the work of
\citet{2012JAVSO..40..295B}.

In this {\it Research Note\/}, we have compiled all previously published
times of mid-eclipse (i.e., from \citet{2012JAVSO..40..295B} and
\citet{2017MNRAS.464..104H}), including the 20 previously unpublished
timings kindly provided by David Sing from his original work
\citep{2007A&A...474..951S}. We have used these timings, along with
seven new eclipse timings that we have recently measured
to update the ephemeris for J0644
and to check the stability of the orbital period.
Of these seven timings, one is based on observations taken
with the Transiting Exoplanet Survey Satellite (TESS).

\section{Observations}

We obtained time-resolved photometric observations of J0644
on six nights in early 2021: Feb 21, Mar 05, 26, 27, and Apr 08, 09 UT.
The observations were arranged to span the eclipse times predicted
by the ephemeris of \citet{2017MNRAS.464..104H}.
The 2021 Feb 21 and Mar 05 UT data were obtained with the
Boyce-Astro Research Initiatives and Education Foundation (BRIEF)
BARO Telescope, which is a
17-in f/6.8 Corrected Dall-Kirkham Astrograph instrument located in
Tierra Del Sol, California and operated remotely.
Observations on the remaining nights
were made with the facilities of the Las Cumbres Observatory (LCO).
The 2021 Mar 27 UT data were obtained with the LCO telescope at
McDonald Observatory, with the rest being obtained at the Haleakala
facility.
All observations consist of
30~s exposures through a Sloan $r$ filter.

In addition to our ground-based eclipse observations,
TESS observed a total
of 86 (mostly sequential) eclipses over the interval between
2019 Dec 25 and 2020 Jan 19 UT. Given that the TESS eclipses were
concentrated over a short time span, with relatively poor
temporal resolution ($\sim2$ min) and signal-to-noise ratio during eclipse,
we have determined a single fiducial eclipse timing for the TESS
observations by fitting a linear function
to the individual times of mid-eclipse and taking
the intercept as the best measure of $T_\mathrm{mid-ecl}$ for this epoch.
Times of mid-eclipse for both the ground-based and TESS observations
were determined by fitting a parabola to the bottom
half (as defined by the inflection points on ingress and egress)
of each of the observed eclipses.

\begin{figure*}
\plotone{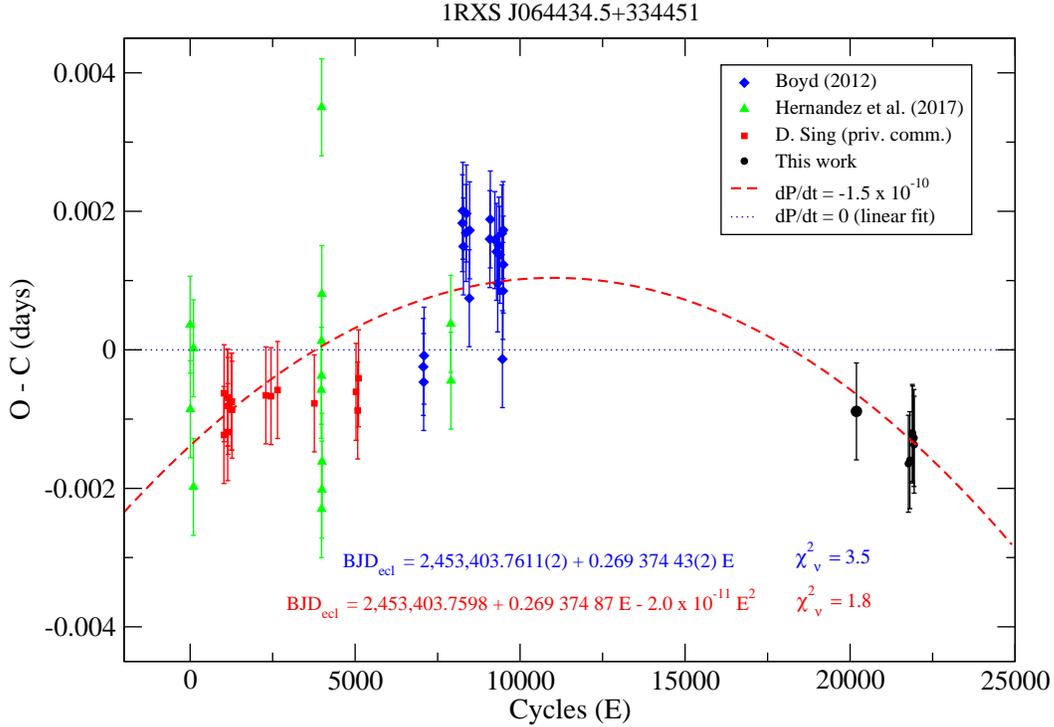}
\caption{The $O-C$ diagram for J0644 showing the residuals from the
best-fitting linear ephemeris given in the text. The residuals have been
fit by a quadratic ephemeris that provides a slightly
better match to the data, and suggests that the orbital period
of J0644 may be decreasing at a rate given by
$\dot P \simeq -1.5\times10^{-10}$.
The sources of data are given in the legend. The larger black filled circle
is based on the mean eclipse timing from the TESS data.
The eclipse timings and $O-C$ values are
available as the Data behind the Figure (see Table A1).
 \label{fig:omc}}
\end{figure*}

With the addition of our new eclipse measurements,
we now have a total of 59
times of mid-eclipse for J0644: 14 from \citet{2017MNRAS.464..104H}, 22 from
\citet{2012JAVSO..40..295B}, 16 previously unpublished timings
provided by D. Sing, along with the 7 from the present study.
A linear least squares fit of all 59 times of mid-eclipse yields
the following ephemeris:
BJD$_\mathrm{ecl} = 2,453,403.7611(2) + 0.269~374~43(2)~\mathrm{E}$.
The individual times of mid-eclipse and their residuals from this
ephemeris are given in the machine-readable table associated with this
{\it Research Note\/}. The residuals are also
plotted in the $O-C$ diagram shown in Figure~\ref{fig:omc}.
The distribution of the residuals suggest that the orbital period
of J0644 may be slowly decreasing. A quadratic fit to all 59 times
of mid-eclipse yield the following ephemeris:
BJD$_\mathrm{ecl} = 2,453,403.7598 + 0.269~374~87~\mathrm{E} - 2.0\times10^{-11}~\mathrm{E}^2$.

In order to estimate reduced $\chi^2$ values of the fits, which are necessary to
assess whether the improved quadratic fit is warranted, uncertainties
must be specified for the individual eclipse timings. To avoid introducing
unknown biases into the analysis, we have assumed that
all eclipse timings are uncertain by 60~s (0.0007~d).
A comparison of the reduced $\chi^2$ values for the two fits
($\chi^2=3.5$ and $\chi^2=1.8$ for the linear and quadratic fits,
respectively) suggests
that neither fit provides a particularly good match to the eclipse timings,
which is not necessarily surprising given that the uncertainties in the eclipse
timings are poorly known and may be somewhat larger than we have assumed
in our analysis.
Nevertheless, if the eclipse timings and their associated 
errors are taken at face value,
it appears that the quadratic fit offers a marginally significant improvement.
In this case, we find that the orbital period of J0644 may be decreasing
at a rate given by: 
$\dot P \simeq -1.5\times10^{-10}$.
Future eclipse observations of J0644 will be required to better establish
the stability of the orbital period.

\begin{acknowledgments}
We thank David Sing for providing unpublished eclipse timings from his
previous work \citep{2007A&A...474..951S},
Pat Boyce of BRIEF for access to the BARO Telescope, Michael Fitzgerald
of the {\it Our Solar Siblings\/} project for assistance with image processing,
and the Las Cumbres Observatory
for a generous allocation of observing time.
\end{acknowledgments}

\vspace{5mm}
\facilities{BARO Telescope, Las Cumbres Observatory}

\bibliography{j0644}{}
\bibliographystyle{aasjournal}

\startlongtable
\begin{deluxetable}{ccrc}
\tablenum{A1}
\tablecolumns{4}
\tablecaption{Eclipse Timings and $O-C$ Residuals}
\tablehead{\colhead{BJD\tablenotemark{a}} & \colhead{Cycle Number} & \colhead{$O-C$} & \\
\colhead{($-2,450,000$)} & \colhead{(E)} & \colhead{(days)} & \colhead{Reference\tablenotemark{b}}
}
\startdata
   3403.7615   &       0   &  $ 0.0004$ & (1) \cr
   3405.6459   &       7   &  $-0.0009$ & (1) \cr
   3430.6966   &     100   &  $-0.0020$ & (1) \cr
   3431.7761   &     104   &  $ 0.0000$ & (1) \cr
   3679.8687   &    1025   &  $-0.0012$ & (2) \cr
   3680.9468   &    1029   &  $-0.0006$ & (2) \cr
   3709.7698   &    1136   &  $-0.0007$ & (2) \cr
   3710.8468   &    1140   &  $-0.0012$ & (2) \cr
   3711.9248   &    1144   &  $-0.0007$ & (2) \cr
   3712.7328   &    1147   &  $-0.0008$ & (2) \cr
   3741.8253   &    1255   &  $-0.0007$ & (2) \cr
   3742.9028   &    1259   &  $-0.0007$ & (2) \cr
   3745.8658   &    1270   &  $-0.0009$ & (2) \cr
   4021.9748   &    2295   &  $-0.0007$ & (2) \cr
   4063.9972   &    2451   &  $-0.0007$ & (2) \cr
   4117.6028   &    2650   &  $-0.0006$ & (2) \cr
   4417.9551   &    3765   &  $-0.0008$ & (2) \cr
   4474.7933   &    3976   &  $-0.0006$ & (1) \cr
   4475.8710   &    3980   &  $-0.0004$ & (1) \cr
   4476.6772   &    3983   &  $-0.0023$ & (1) \cr
   4476.9490   &    3984   &  $ 0.0001$ & (1) \cr
   4477.7605   &    3987   &  $ 0.0035$ & (1) \cr
   4478.8353   &    3991   &  $ 0.0008$ & (1) \cr
   4479.6406   &    3994   &  $-0.0020$ & (1) \cr
   4480.7185   &    3998   &  $-0.0016$ & (1) \cr
   4757.9058   &    5027   &  $-0.0006$ & (2) \cr
   4771.9130   &    5079   &  $-0.0009$ & (2) \cr
   4778.9172   &    5105   &  $-0.0004$ & (2) \cr
   5307.4300   &    7067   &  $-0.0002$ & (3) \cr
   5310.3929   &    7078   &  $-0.0005$ & (3) \cr
   5313.3564   &    7089   &  $-0.0001$ & (3) \cr
   5531.0114   &    7897   &  $ 0.0004$ & (1) \cr
   5532.8962   &    7904   &  $-0.0004$ & (1) \cr
   5627.4489   &    8255   &  $ 0.0018$ & (3) \cr
   5629.3347   &    8262   &  $ 0.0020$ & (3) \cr
   5634.4523   &    8281   &  $ 0.0015$ & (3) \cr
   5655.4637   &    8359   &  $ 0.0017$ & (3) \cr
   5658.4271   &    8370   &  $ 0.0020$ & (3) \cr
   5682.4002   &    8459   &  $ 0.0007$ & (3) \cr
   5685.3643   &    8470   &  $ 0.0017$ & (3) \cr
   5850.4907   &    9083   &  $ 0.0016$ & (3) \cr
   5854.5316   &    9098   &  $ 0.0019$ & (3) \cr
   5891.4356   &    9235   &  $ 0.0016$ & (3) \cr
   5905.4429   &    9287   &  $ 0.0014$ & (3) \cr
   5914.3318   &    9320   &  $ 0.0010$ & (3) \cr
   5924.2992   &    9357   &  $ 0.0015$ & (3) \cr
   5932.3803   &    9387   &  $ 0.0014$ & (3) \cr
   5949.3512   &    9450   &  $ 0.0017$ & (3) \cr
   5953.3900   &    9465   &  $-0.0001$ & (3) \cr
   5957.4316   &    9480   &  $ 0.0008$ & (3) \cr
   5959.3181   &    9487   &  $ 0.0017$ & (3) \cr
   5960.3951   &    9491   &  $ 0.0012$ & (3) \cr
   8842.6994   &   20191   &  $-0.0009$ & (4) \cr
   9266.6940   &   21765   &  $-0.0016$ & (4) \cr
   9278.8159   &   21810   &  $-0.0016$ & (4) \cr
   9299.8275   &   21888   &  $-0.0012$ & (4) \cr
   9300.6356   &   21891   &  $-0.0012$ & (4) \cr
   9312.7574   &   21936   &  $-0.0013$ & (4) \cr
   9313.8348   &   21940   &  $-0.0014$ & (4) \cr
\enddata
\tablecomments{This table is available in machine-readable format in the {\it AAS Research Notes.}}
\tablenotetext{a}{BJD mid-eclipse times are computed with respect to the Barycentric Dynamical Time (TDB) standard.}
\tablenotetext{b}{(1) \citet{2017MNRAS.464..104H}; (2) D. Sing (private communication); (3) \citet{2012JAVSO..40..295B}; (4) This work.}
\end{deluxetable}

\end{document}